\documentclass{PoS}

\title{Radio and gamma-ray emission in faint BL Lacs.}

\ShortTitle{Radio and gamma-ray emission in faint BL Lacs.}

\author{\speaker{E.Liuzzo}\\%\thanks{A footnote may follow.}\\
        Istituto di Radioastronomia-INAF-Bologna (Italy)\\
        E-mail: \email{liuzzo@ira.inaf.it}}

\author{B. Boccardi\\
       Max Planck Institute for Radioastronomy- Bonn (Germany)\\
        E-mail: \email{bboccardi@mpifr-bonn.mpg.de
}}

\author{M.Giroletti\\
       Istituto di Radioastronomia-INAF-Bologna (Italy)\\
        E-mail: \email{giroletti@ira.inaf.it}}

\author{G.Giovannini\\
       Universit\`a degli Studi di Bologna and Istituto di Radioastronomia-INAF-Bologna (Italy)\\
        E-mail: \email{ggiovann@ira.inaf.it}}

\abstract{The advent of {\it Fermi} is changing our understanding on the radio and $\gamma$-ray emission in Active Galactic Nuclei. In fact, contrary to previous campaigns, {\it Fermi} mission reveals that BL Lac objects are the most abundant emitters in $\gamma$-ray band. However, since they are relatively weak sources, most of their parsec scale structure as their multifrequency properties are poorly understood and/or not systematically investigated.
   Our main goal is to analyse, using a multiwavelength approach, the nuclear properties of an homogeneous sample of 42 faint BL Lacs, selected, for the first time in literature, with no constraint on their radio and $\gamma$-ray flux densities/emission. We began asking and obtaining new VLBA observations at 8 and 15 GHz for the whole sample. We derived fundamental parameters as radio flux densities, spectral index information, and parsec scale structure. Moreover, we investigated their $\gamma$-ray emission properties using the 2LAT Fermi results.
Here, we report our preliminary results on the radio and gamma-ray properties of this sample of faint BL Lacs. In the next future, we will complete the multiwavelength analysis.}

\FullConference{11th European VLBI Network Symposium \& Users Meeting,\\
		October 9-12, 2012\\
		Bordeaux, France}

\begin{document}

\section{Introduction} 

Since the measurement from  Energetic Gamma-Ray Experiment Telescope (EGRET) on the Compton Gamma-Ray Observatory (CGRO), it is well know that at the $\gamma$-ray sky contribute the Galactic plane emission, pulsars, and blazars (\cite{ha}). Among the latter class of objects, Flat Spectrum Radio Quasars (FSRQs) were most numerous than the BL Lacs, being the 77$\%$ of the high confidence blazar associations. 

With the advent of the {\it Fermi} mission, studies of the multiwavelength properties of a large number of $\gamma$-ray sources could be performed thank to its unprecedented sensitivity. Moreover, the positional accuracy of the Large Area Telescope (LAT) on board significantly improved  the localization of the $\gamma$-ray sources with respect to the past $\gamma$-ray campaigns. In contrast with the previous EGRET results, the LAT has shown that the BL Lacs, and not the FSRQs, are now the most common $\gamma$-ray emitters (1LAC \cite{abd}, 2LAC \cite{ac}). Moreover, the gamma-ray properties of the two blazar sub-classes are markedly distinct, e.g. in the average photon indices, redshift and flux density distributions, etc.  Big questions are nevertheless remaining open, as the $\gamma$-ray origin in relativistic shocks, the distance of the main energy dissipation site from the nucleus, the relation between the $\gamma$-ray and radio emission.

High resolution and VLBI (Very Long Baseline Interferometer) campaigns seem to be one of the most incisive observational tools to address the {\it Fermi} era. From a detailed review from the literature (\cite{ro}), it is evident that the parsec scale properties of BL Lacs are poorly studied in a systematic way, as VLBI surveys involved in are generally affected by small numbers,  high flux density limits, incompleteness, and selection effects (\cite{wu, gir04, gir06, re, ca}). In particular, many $\gamma$-ray BL Lacs are
high-synchrotron-peaked (HSP) sources, discovered at X-rays and generally faint radio sources, and they have been only seldom studied with VLBI.

To improve our knowledge of their nuclear region, we started a project to perform a systematic analysis defining an homogeneous sample of BL Lacs, not selected on the base of their radio flux densities and their $\gamma$-ray emission. 
We selected a sample of BL Lacs from the ASDC Catalog of known blazars (Roma-BZCat, \cite{ma}) with the only following two constraints: 1) a measured redshift z $<$ 0.2, 2) BL Lacs located within the sky area covered by the Sloan Digital Sky Survey (SDSS, \cite{aba}). 
The redshift constraint allows us to define a not biased sample from the optical point of view. We assume that for all BL Lacs at z$\leq$0.2 it is possible to have a redshift estimation. Moreover, these two criteria allow us 1) a good linear resolution (1 pc $\sim$ 0.5 mas at z = 0.1) to investigate also the least powerful sources, such as the weak population of HSP BL Lacs; 2) a good characterization not only of the optical properties, but also of their extended radio characteristics (as the FIRST covered the same SDSS field), and 3) to have information on the X-ray emission (from ROSAT data, see also \cite{ma}).

The total number of BL Lacs in this sample is 42. In Table \ref{tab_intro} we report the main properties of each source. Looking at the distribution of their NVSS flux densities, we point out that our BL Lacs sample is still unexplored by two of the most complete previous VLBI surveys, the MOJAVE (with correlated flux S $>$ 1.5 Jy (\cite{li}) and VIPS (S $ >$ 85 mJy, \cite{he}), as they are flux limited surveys.

\begin{table}
\caption{{\bf Main properties} of the sample}\label{tab_intro}
\centering
\footnotesize
\begin{tabular}{|c|c|l|r|r|r|r|}
\hline
Name & z &  RA(J2000)& Dec(J2000)&$M_R$ & $S_{NVSS}$& $S_{FIRST}$  \\
 & &h m s               & d m s              & & (mJy) & (mJy)\\
\hline
\hline
J0751+1730 & 0.185 & 07 51 25.08 & +17 30 51.1 & 18.2 & 9.72  &10.96\\
J0751+2913 & 0.194 & 07 51 09.57 & +29 13 35.5 & 16.9 & 12.38 & 8.92\\
J0753+2921 & 0.161 & 07 53 24.61 & +29 21 21.9 & 15.7 & 3.96  & 4.49 \\
J0754+3910 &0.096&07 54 37.08   & +39 10 47.6&12.8&57.8&49.26\\
J0809+3455 &0.083&08 09 38.87    & +34 55 37.1&12.6&227.43&169.12\\
J0809+5218 & 0.138&08 09 49.19     & +52 18 58.2   &  14.6 & 183.82 & 187.05\\
J0810+4911 & 0.115 & 08 10 54.60  & +49 11 03.7 & 13.5 & 10.76 & 10.91\\
J0847+1133 &0.199&08 47 12.94     & +11 33 50.1&16.6&32.98&33.66\\
J0850+3455 &0.145&08 50 36.18      & +34 55 22.8&14.3&34.51&30.86\\
J0903+4055 & 0.188 & 09 03 14.71  & +40 55 59.9 & 15.8 & 35.75 & 29.75\\
J0916+5238 & 0.190 &09 16 51.94     & +52 38 28.5 &15.0 & 88.41& 108.93\\
J0930+4950 & 0.187 &09 30 37.57     & +49 50 25.6   & 17.3 &21.33 &15.08 \\
J1012+3932 & 0.171&10 12 58.37  & +39 32 39.0& 16.0& 19.0 & 20.11 \\ 
J1022+5124 & 0.142 & 10 22 12.62  & +51 24 00.3 &16.7& 5.59 & 2.69 \\
J1053+4929 & 0.140 &10 53 44.10    & +49 29 55.9   &13.8 &65.45 & 62.61 \\
J1058+5628 & 0.143 &10 58 37.73    & +56 28 11.2 &  14.0& 229.48 &219.45 \\ 
J1120+4212 & 0.124 & 11 20 48.06     & +42 12 12.5   & 16.9& 23.54 & 24.56\\
J1136+6737 & 0.136 & 11 36 30.09    & +67 37 04.3   & 15.3& 45.15 &- \\
J1145-0340 & 0.167 & 11 45 35.11   & - 03 40 01.7  &16.2 & 18.65 &10.48\\
J1156+4238 & 0.172 & 11 56 46.56   & +42 38 07.4  &15.6 & 15.64 &14.38 \\
J1201-0007 &0.165&12 01 06.22      & -00 07 01.8&15.7&69.51&67.57\\
J1201-0011 &0.164&12 01 43.66      & -00 11 14.0&16.7&27.98&23.47\\
J1215+0732 & 0.136 & 12 15 10.97   & +07 32 04.7 &14.8 & 138.81 & 81.80\\
J1217+3007 & 0.130 & 12 17 52.08      & +30 07 00.5&14.5& 587.82& 466.45\\
J1221+0821 &0.132&12 21 32.06      & +08 21 44.1&16.3&178.36&162.53\\
J1221+2813 & 0.102 & 12 21 31.69    & +28 13 58.5 &14.3 & 738.97 &921.26\\
J1221+3010 & 0.182 &12 21 21.94      & +30 10 37.1& 15.7& 72.01&62.49\\
J1231+6414 & 0.163 &12 31 31.40     & +64 14 18.3 & 14.3 & 59.31 &- \\ 
J1253+0326&0.066&12 53 47.03      & +03 26 30.4&12.7&107.35&79.21 \\
J1257+2412 & 0.141 &12 57 31.93     & +24 12 40.1  & 15.7 & 13.07 &10.32 \\ 
J1341+3959 &0.172&13 41 05.11     &+39 59 45.4 &14.9&85.63&57.85\\
J1419+5423 & 0.153 &14 19 46.60      & +54 23 14.8&13.8 & 818.16&581.55\\  
J1427+3908 & 0.165 &14 27 45.92     & +39 08 32.3 & 18.0 & 6.96 &4.79\\
J1427+5409 & 0.106 &14 27 30.28      & +54 09 23.7& 11.8 & 44.76 &29.79 \\
J1428+4240 &0.129&14 28 32.62       &+ 42 40 21.2&14.4&57.52&42.72\\
J1436+5639 & 0.150 &14 36 57.72      & +56 39 24.9&17.6 &20.71 & 17.11\\ 
J1442+1200 &0.163&14 42 48.27        &+12 00 40.3&15.2&67.95&69.97\\
J1510+3335 & 0.114 & 15 10 41.18     & +33 35 04.5  & 15.1 & 7.36 &4.10\\ 
J1516+2918 &0.130&15 16 41.60         &+29 18 09.5&14.6&136.51&73.96\\
J1534+3715 & 0.143 &15 34 47.21     & +37 15 54.6  &  16.3 & 20.96 &21.57\\
J1604+3345 & 0.177 &16 04 46.52    & +33 45 21.8  & 18.0 & 7.09&5.84 \\
J1647+2909 & 0.132 & 16 47 26.88     &+29 09 49.6& 13.4 & 394.72 &275.79 \\
\hline
\multicolumn{7}{l}{Col.1: Name of source; Col. 2 redshift;Col.s 3 - 4: coordinates (J2000);} \\
\multicolumn{7}{l}{Col. 5 optical Magnitude in band R;}\\
\multicolumn{7}{l}{Col.s 6 - 7 total radio power at 1.4 GHz from NVSS and from FIRST data.}
\end{tabular}
\end{table}
\clearpage

\section{New radio data.} 

\begin{figure}[t!]
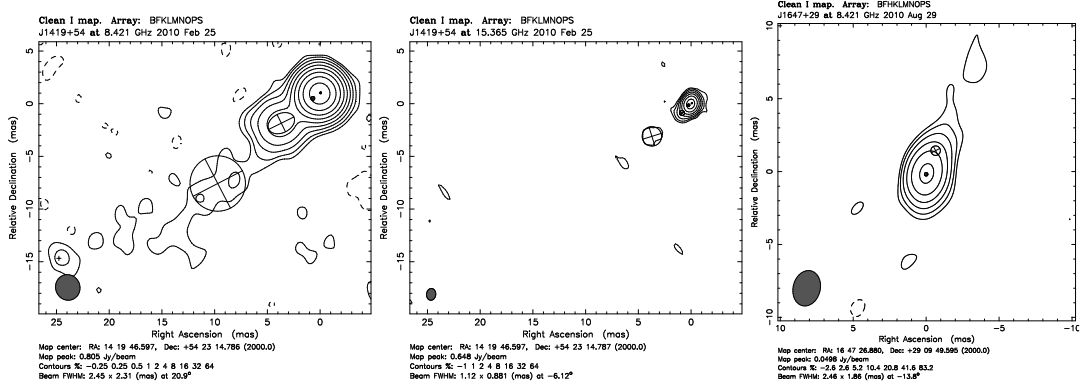

\centering
\includegraphics[width=0.32\textwidth]{fig1.ps}
\includegraphics[width=0.32\textwidth]{fig2.ps}
\includegraphics[width=0.29\textwidth]{fig3.ps}
\caption{Contours images of J1419+5423 in X ({\it left panel}) and U bands ({\it central panel}). In the {\it right panel}, we report contours image of J1647+2909 in X band (the source is point like at 15 GHz).}  \label{fig_es}
\end{figure}

We asked and obtained new Very Long Baseline Array (VLBA) observations for all sources of the sample and we present here
our first results. Each target was observed at 8 and 15 GHz with the aim of obtaining simultaneous spectral information. The observing time was about one hour per source, with roughly a 1:3 ratio between the low and high frequency total integration time. Targets weaker than 30 mJy at 8 GHz and 50 mJy at 15 GHz have been observed in phase referencing mode. Indeed, most sources had not been observed before and the phase referencing technique has also provided the possibility to obtain absolute coordinates for them. The observations were carried out using 8 or 9 VLBA telescopes at a 256 Mbps recording rate; the data quality is overall good and permits us to achieve a noise levels of $\sim$ 0.2 mJy/beam, i.e. comparable to the theoretical
one. The restoring beam is the typical one of the VLBA at our frequencies, i.e. $\sim$ 1.2 $\times$ 1.8 mas at 8 GHz and $\sim$ 0.6 $\times$ 0.9 mas at 15 GHz, for sources at intermediate declination and with natural weights. 
Figure \ref{fig_es} reports contours images of two among our detected sources. 

The detection rate is 67$\%$ at 8 GHz and 57$\%$ at 15 GHz. Point-like morphologies are present in the 40\% of the detected sources at 8 GHz, and in 70\% of the revealed targets at 15 GHz. One-sided structures are found in 64\% of detected sources at 8 GHz, and in 7 objects at 15 GHz (30\% of the detected sources at 15 GHz). We note that, among the 6 BL Lacs having one sided morphology at the both frequencies, 5 are the most luminous BL Lacs at mas scale.  The sample is mostly composed of quite faint objects, even if there are some powerful radio sources, like J1217+3007 or J1419+5423. In fact, the largest fraction ($70\%$) of detected sources has flux density in the range 10-40 mJy at 8 GHz, and between 4 and 20 mJy at 15 GHz. We estimated the spectral index distributions between 8 and 15 GHz VLBA data ($\alpha_{15}^{8}$): the fraction of flat spectral indices is predominant as expected for
BL Lac objects. Therefore, there is also a considerable and unexpected number of sources with a steep spectrum $\alpha\geq$1. An unexpected number of objects are also found with low source compactness (SC, which is defined as the ratio of the 8 GHz VLBA and the NVSS total powers). 

Looking at pc and kpc scale properties, BL Lacs could be separate in four class of objects:\\
1) {\it Compact (C) sources:}  3 targets (J1058+5628, J1120+4212, J1419+5423) have high source compactness (SC $\geq$0.72) sources, with similar mas and arcsec flux densities, and flat $\alpha_{15}^{8}$ ($\sim$0.20-0.25). \\
2) {\it Partially resolved (pR) objects:} in $\sim$45\% of targets, the VLBA core is clearly visible at both frequencies. However, the VLBA core flux densities are 70\% of these detected in NVSS maps. This suggests the presence of a sub-kpc structure that we are not able to image in our mas scale observations. Moreover, their $\alpha_{15}^{8}$ are flat or moderately steep. \\
3) {\it Resolved (R) sources:} In $\sim$30\% of objects, the VLBA fluxes are a small fraction ($\leq$40\%) of the kpc scale ones. These BL Lacs show low source compactness (SC $<$0.25) and/or steep spectral index.\\
4) {\it Undeterminate (U) objects:} there are few sources that are not detected in our VLBA images, and they have low arcsec flux densities ($<$10 mJy). Deeper observations would be necessary to better investigate their parsec scale emission.

\clearpage

\section{Gamma-ray properties.} 

We search for gamma-ray counterparts of our sample BL Lacs: 14/42 of the BL Lacs show high energy emission in the 2LAC (Ackermann et al. 2011). We named them LAT BL Lacs. Among these LAT BL Lacs, 3 are C sources (J1058+5628, J1120+4212, J1419+5423) and 11 are pR objects. No R objects are present. Correlations for LAT BL Lacs are found between the 8 GHz VLBA radio S$_{VLBA, 8 GHz}$  and gamma-ray fluxes, and between S$_{VLBA, 8 GHz}$ and photon indices, while no correlation is observed between X and gamma-ray fluxes. This is in agreement with a Synchrotron Self Compton scenario for the production of gamma-ray emission in these BL Lacs.

Comparing results in radio band for LAT and non LAT BL Lacs, we found that the LAT BL Lacs are the most luminous at pc scale and they are the all sources in the sample which are one-sided at both frequencies. In particular, all LAT BL Lacs with Log P$_{VLBA, 8 GHz}$(W/Hz)$\geq$24.5 present resolved morphologies both at 8 and 15 GHz. On the other hand, objects with Log P$_{VLBA, 8 GHz}\leq$23.5 do not show gamma -ray emission, as sources with source compactness lower than 0.28. Moreover, the kiloparsec scale radio emission seems not to be important for the gamma-ray emission, in the sense that,e.g. J1647+2909 has S$_{NVSS}\sim$395 mJy but it does not emit in the gamma-rays, while J1120+4212 has S$_{NVSS}\sim$24 mJy and it is a LAT BL Lac.

\begin{figure} [t!]
\centering
\includegraphics[width=0.5\textwidth]{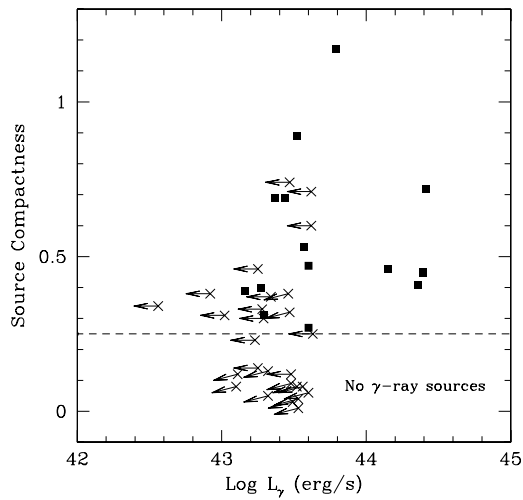}
\caption{{\bf Log L$_{\gamma}$} vs {\bf Source compactness} for LAT BL Lacs (filled squares) and non LAT BL Lacs (crosses). Arrows indicate upper limits}.\label{pgsc}
\end{figure}

Finally, gamma-ray luminosities were calculated for the whole sample. Upper limits were considered for non-detected sources, both in the radio and in the gamma-ray band. SC  upper limits are calculated from the upper limit flux density (3$\sigma$) of our VLBA images. In the gamma-rays, we attributed to the non-detected sources a photon flux equal to the minimum flux in the sample, i.e that of J1534+3717 (4.49$\times$10$^{-10}$ ph cm$^{2}$ s$^{-1}$), and the sample average photon index value ($\Gamma$=1.78). We plot our results in Fig. \ref{pgsc}: the main note is that among the gamma-ray sources, there are not sources with SC $<$ 0.28.

Concluding, our radio and gamma-ray study of this sample of faint BL Lacs suggests that the source compactness and the parsec scale flux density are the most relevant parameters determining their high energy emission. Next steps will be the analysis of the remaining multiwavelength data, characterizing the nuclear properties of these objects on the whole spectrum and investigating better these differences among LAT and non-LAT BL Lacs.

\acknowledgments

We thank the organizers of a very interesting meeting. This work was supported by contributions of European Union, Valle DAosta Region and the Italian
Minister for Work and Welfare.

\end{document}